\documentclass[hyper]{JHEP}
\pdfoutput=1
\usepackage{fixltx2e}
\usepackage{float}
\usepackage{graphicx,amssymb,bm,latexsym,amsmath}
\usepackage{subfigure,float,psfrag,rotating}
\usepackage{epstopdf}
\usepackage{caption}
\newcommand{\sn}{{\rm sn}}
\newcommand{\cn}{{\rm cn}}
\newcommand{\dn}{{\rm dn}}

\title{Perturbations of spiky strings in $AdS_3$}
\author{Soumya Bhattacharya, Sayan Kar and  Kamal L. Panigrahi \\
Department of Physics and Centre for Theoretical Studies,\\Indian Institute of Technology Kharagpur,\\
Kharagpur-721 302, India\\
Email: \email{soumya557@cts.iitkgp.ernet.in, sayan,panigrahi@phy.iitkgp.ernet.in}}

\abstract{Perturbations of a class of semiclassical spiky strings
in three dimensional Anti-de Sitter (AdS) spacetime, are investigated 
using the well-known Jacobi
equations for small, normal deformations of an embedded timelike
surface. We show that the equation for the perturbation scalar
which governs the behaviour of such small deformations, is a
special case of the well-known Darboux-Treibich-Verdier (DTV) equation. 
The eigenvalues and eigensolutions of the DTV equation for our case
are obtained
by solving certain continued fractions numerically. These solutions are
thereafter utilised to further
demonstrate that there do exist finite perturbations of the
AdS spiky strings. Our results therefore establish that the 
spiky string configurations in $AdS_3$ are indeed stable against small fluctuations.
Comments on future possibilities of work are included in conclusion.}


\begin{document}

\section{Introduction}
In the early days of string theory,
rigidly rotating strings arose in the work of Burden and Tassie
\cite{BurdenTassie1: 1982, BurdenTassie2: 1982,BurdenTassie3: 1984} primarily as stringy 
models for exotic mesons and glue balls. Gravitational radiation from
this particular class of strings (viewed as cosmic strings) 
was discussed in \cite{Burden: 1985}. Later, Embacher
\cite{Embacher: 1992}  obtained the complete class of such
solutions in flat spacetime, with rigid rotation about a
fixed axis.  Further work  can be found in \cite{rigid:1,rigid:2,rigid:3,Burden: 2008, ogawa:2008}.

\noindent In its modern avatar, such rigidly rotating strings have been
renamed as {\em spiky strings}. Their importance and relevance today 
is largely in the context of  the AdS/CFT correspondence
\cite{Maldacena:1997re}. More than a decade ago, the spiky string solutions
appeared, through the seminal work by Kruczenski
\cite{Kruczenski:2004wg}, as a potential gravity dual to higher
twist operators in string theory. In string theory and in AdS/CFT,
the solutions of interest are closed strings with
spikes (in cosmic string nomenclature these are called `cusps').
The semiclassical aspect of these solutions arise through
evaluating their energies, angular momenta and finding relations between them \cite{Gubser:2002tv}.  

\noindent 
In providing meaningful input towards the realization of the gauge-gravity duality, the
emergence of integrability on both sides have been quite useful and important. 
Classical and quantum integrability of the ${\cal N}$ = 4 Supersymmetric Yang-Mills (SYM) theory in the
planar limit is mainly useful for the remarkable advances in understanding the theory \cite{Minahan:2002ve, dolan, nappi}. The
nonlocal conserved charges found on the string side \cite{Mandal:2002fs, Bena:2003wd} appear to have a counterpart in planar
gauge theory at weak coupling within the spin-chain formulation for the dilatation operator.
%\cite{nappi}. 
The integrability of the ${\cal N} $ = 4 SYM
theory should not only have important consequences for the spectrum of anomalous dimensions
of gauge-invariant single trace operators but also for other observables, e.g. the Wilson
loops. One such class of string states in the string theory side are the so called spiky strings, 
which have been shown to be dual to higher twist operators in ${\cal N}$ = 4 SYM theory with 
each spike corresponding to a particle on the field theory side of the
correspondence. Large angular momentum is
provided by a large number of covariant derivatives acting on the fields which produce the 
above-mentioned particles. The total number of derivatives is distributed equally amongst the 
fields for these solutions. A large class of such spiky strings in various asymptotically AdS and non-AdS  backgrounds have
been studied, for example, in
\cite{Kruczenski:2004wg},\cite{Frolov:2003qc},\cite{Ishizeki:2007we},\cite{Ishizeki:2008tx},\cite{Biswas:2012wu},\cite{Banerjee:2014gga},\cite{Banerjee:2015nha}. 

\
\noindent Given the above-stated importance of the spiky strings, it is  
worthwhile to look at the geometric
properties of such string configurations from the world sheet view
point. To this end, following our earlier work  \cite{bkp:2017} we study normal deformations (linearized) about the
classical solutions in $AdS_3$. Earlier work in \cite{Frolov:2002av} dealt with
computation of quantum corrections to the energy spectrum, 
by expanding the supersymmetric action to quadratic order in
fluctuations about the classical solution. The linearized perturbations of semiclassical
strings are extremely instructive in matching the duality beyond the leading order 
classical solutions.  The main motivations behind the perturbative solutions are multi-fold. On one hand, it helps in studying the stability properties
of the string dynamics of the closed strings, and finding the quantum string corrections to 
the Wilson loop expectation value for the open string solutions and on the other hand, it helps us in determining the physical properties of topological defects.

\noindent Motivated by various studies on general classes of rotating strings in $AdS_5 \times S^5$, in this paper, we investigate the classical stability
of the spiky strings in $AdS_3$ using the well-known Jacobi equations
\cite{garriga: 1993},\cite{guven: 1993},\cite{frolov: 1994},
\cite{capovilla: 1995} which govern normal deformations of an
embedded surface. Apart from \cite{bkp:2017}, related recent work on such perturbations can be found
in \cite{forini:2015}.
The rest of this article is organised as follows. In Section 2, we
briefly summarize the two different embeddings
of the worldsheet and the spiky string solution in $AdS_3$. 
Section 3 is devoted to a study of the perturbation
equation which turns out to be a special case of the Darboux-Treibich-Verdier equation in mathematical physics. We also discuss, in Section 3, the
numerical evaluation of the eigenvalues and eigenfunctions, obtained by solving
an infinite continued fraction numerically.  Plots of the
perturbations $\delta x^i$ are shown for various eigenvalues
and the issue of stability is addressed.
Our final concluding remarks appear in Section 4.

\section{Spiky strings in $AdS_3$:
Kruczenski and Jevicki-Jin embeddings}

\noindent The bosonic string  worldsheet embedded in a 
$N$ dimensional curved spacetime with background metric
functions $g_{ij}(x)$, is described by the well-known Nambu-Goto action given as,
\begin{equation}
S =-T \int d\tau \,\, d\sigma \sqrt{-\gamma} = - T \int  d\tau \,\, 
d\sigma \sqrt{ (\dot{x}\cdot x')^2-\dot{x}^2 {x'}^2} \ .
\label{ac1}
\end{equation}
Here $\gamma$ denotes the determinant of the induced metric $\gamma_{ab}=g_{ij}
\partial_a x^i \partial_b x^j$ ($a,b=\sigma,\tau$). 
The $x^i(\tau,\sigma)$ are functions which describe the
profile of the string worldsheet, as embedded in the target spacetime.
$T$ denotes the string tension. We have used the notation: $\dot x = \partial_{\tau} x$,\,
$ x' = \partial_{\sigma} 
x$, \, $(\dot x \cdot x') = g_{ij}{\dot x}^{i} \, {x'}^{j}$,\, 
${\dot x}^2=g_{ij}{\dot x}^i \,{\dot x}^j$ and
${x'}^2=g_{ij}{ x'}^i \,{x'}^j$.
\noindent Let us now turn to spiky strings in AdS spacetime in three
dimensions. The AdS line element is given as
\begin{equation}
ds^2 = -\cosh^2\rho\,\,dt^2 +d\rho^2 + \sinh^2 \rho \,\,d\theta^2 \ .
\end{equation}
%Kruczenski's choice of embedding is: 
We use the following embedding
\begin{equation}
 t= \tau , \ \ \theta = \omega\tau + \sigma, \ \ \rho = \rho(\sigma) \ .
\end{equation}

\noindent Using the equations of motion from the action (\ref{ac1}) one can get an equation of $\rho(\sigma)$ given as,
\begin{equation}
\rho' = \frac{1}{2}\frac{\sinh2\rho}{\sinh2\rho_0}\frac{\sqrt{\sinh^22\rho-\sinh^22\rho_0}}{\sqrt{\cosh^2\rho-\omega^2\sinh^2\rho}} \, \, ,
\end{equation}
where $\rho_0$ is the integration constant. From the expression of $\rho'$ it is clear that  $\rho$ varies
from a minimum value $\rho_0$ to a maximum value $\rho_1=\mbox{arccoth}\,\omega$.  At $\rho=\rho_1$, $\rho'$ diverges
indicating the presence of a spike and at $\rho=\rho_0$, $\rho'$ vanishes, indicating the bottom of the valley between spikes. To get a solution 
with $n$ spikes one has to glue $2n$ of the arc segments and the angle between cusp and valley is $\frac{2 \pi}{2n}$. Figures 1 and 2 show such spiky string configurations as embedded in
an Euclidean space, with ten and three spikes respectively \cite{Kruczenski:2004wg}.
\begin{figure}[h]\label{f8}
 \begin{minipage}{18pc}
 \includegraphics[width=22pc]{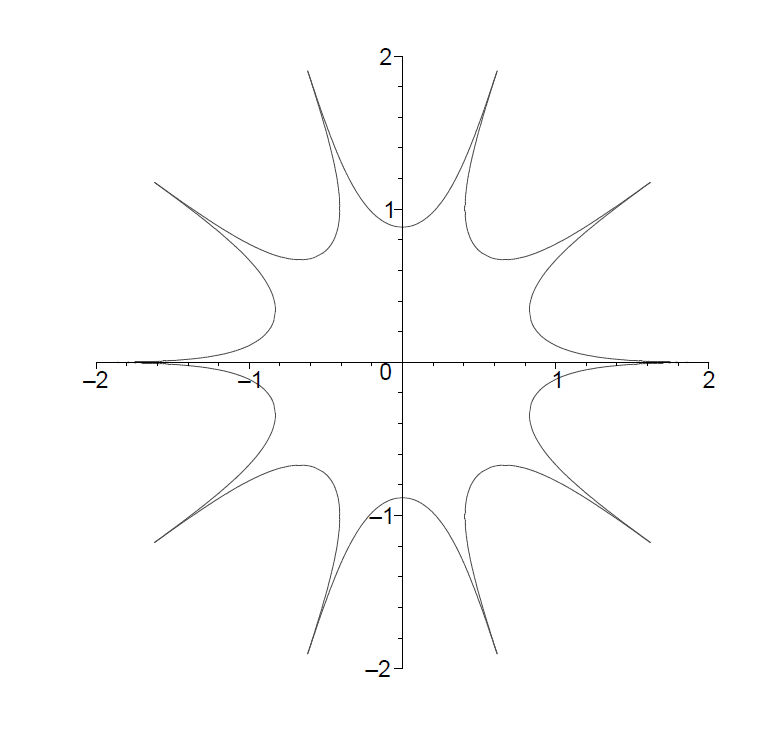}
 \caption{The Kruczenski spiky string in ($\rho,\theta$) plane for $n=10$ spikes with $\rho_1 = 2.0 ~\& ~\rho_0 = 0.88266$ .}
\end{minipage}
%\end{figure}
%\begin{figure}[h]\label{f8}
\hspace{0.5in}
 \begin{minipage}{18pc}
 \includegraphics[width=22pc]{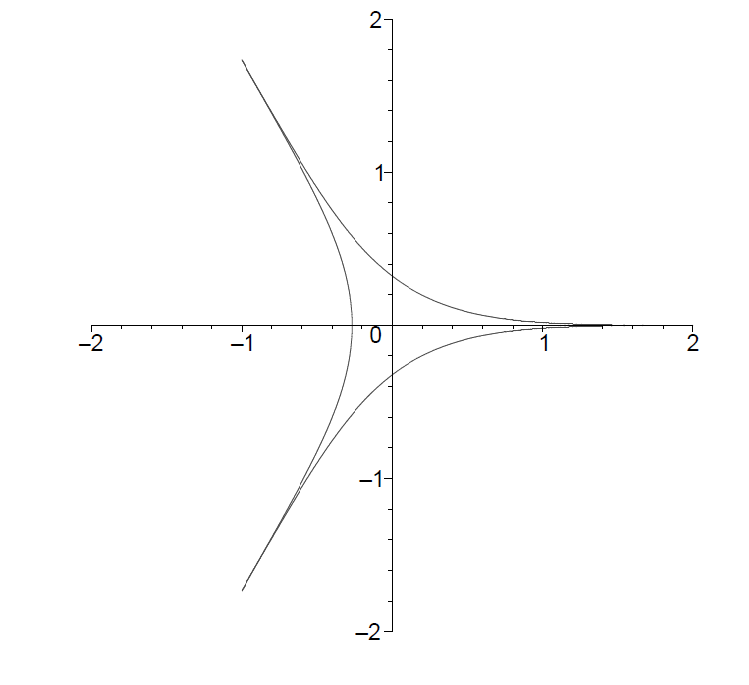}
 \caption{The Kruczenski spiky string in ($\rho,\theta$) plane for $n=3$ spikes with $\rho_1 = 2.0 ~\& ~\rho_0 = 0.2688 $ .}
\end{minipage}
\end{figure}

\noindent In conformal gauge, 
%where the induced metric is diagonal, 
we have
\begin{equation}
g_{ij}({\dot x}^{i}{\dot x}^{j}+{x'}^{i}{x'}^{j}) = 0, \hspace{0.3in} g_{ij}{\dot x}^{i}{x'}^{j} = 0 \ .
\label{ee2}
\end{equation}

\noindent In such a gauge, the string equations of motion 
obtained by varying the action with respect to $x^i$, take the form:
\begin{equation}
{\ddot x}^{i} - {x^{i ''}} + \Gamma^{i}_{jk} \left
( {\dot x}^{j} {\dot x}^{k} - {x}^{j \prime} {x}^{k\prime}
\right )  = 0 \ .
\label{ee1}
\end{equation}
We may also choose the following embedding due to 
Jevicki and Jin \cite{jevicki:2008}:
\begin{equation}
t= \tau + f(\sigma) , \hspace{0.1in} \rho=
\rho(\sigma) , \hspace{0.1in} \theta= \omega \tau +
g(\sigma) \ .
\end{equation}
As we shall see, this embedding will be useful in our calculations later.

\noindent With the Jevicki-Jin ansatz mentioned above, we find that
the tangent vector to the worldsheet is given as:
\begin{equation}
e^i_\tau = \left (1,0,\omega\right ) \hspace{0.2in};\hspace{0.2in}
e^i_\sigma = \left ( f', \rho',g'\right ) \ .
\end{equation}
The equations of motion and the Virasoro constraints turn out to be
\begin{eqnarray}
f'(\sigma) = \frac{\omega \sinh 2\rho_0}{2\cosh^2\rho} \hspace{0.2in}
; \hspace{0.2in} g'(\sigma) = \frac{\sinh 2\rho_0}{2\sinh^2\rho} \\
{\rho'}^2 (\sigma) = \frac{\left (\cosh^2\rho -\omega^2\sinh^2\rho\right )
\left (\sinh^2 2\rho -\sinh^2 2\rho_0\right )}{\sinh^2 2\rho} \ .
\end{eqnarray}
The solution of the $\rho$ equation is given in terms of Jacobian elliptic functions
\begin{equation}
\rho(\sigma) =\frac{1}{2} \cosh^{-1} \left (\cosh 2\rho_1 \,\,\cn^2(u,k)+
\cosh 2\rho_0\,\, \sn^2 (u,k)\right ) \, \, ,
\end{equation}
here $\sn(u,k)$ and $\cn(u,k)$ are  the 
Jacobi elliptic functions with modular parameter $k^2$ and $u$ \& $k$ are defined as
\begin{eqnarray}
u= \sqrt{\frac{\cosh 2\rho_1 +\cosh 2\rho_0}{\cosh 2\rho_1-1}}\,\, \sigma = \zeta \sigma\, \, ,
\label{z1}
\\
k= \sqrt{\frac{\cosh 2\rho_1-\cosh 2\rho_0}{\cosh 2\rho_1 +\cosh 2\rho_0}} \ .
\end{eqnarray}
It is possible to write down $f$ and $g$ as well but we do not need them
here. Explicit expressions are available in \cite{jevicki:2008}. We now turn towards
examining the stability of the string configurations by studying their normal
deformations.

\section{Perturbations and stability of spiky strings in $AdS_3$}
Before investigating the perturbation equations for
our specific solutions, let us briefly recall the well-known Jacobi
equations which deal with
perturbations of extremal worldsheets.

\subsection{Jacobi equations for extremal surfaces}
Given that $x^i(\tau,\sigma)$ are the embedding functions and
$g_{ij}$ the background metric, the tangent vectors to the worldsheet are
\begin{equation}
e^i_{\tau} = \partial_\tau x^i , \hspace{0.2in}
e^i_{\sigma} = \partial_\sigma x^i .
\end{equation}
Thus, the induced line element turns out to be,
\begin{equation}
\gamma_{ab} = g_{ij} e^i_a e^j_b \ ,
\end{equation}
where the $a,b...$ denote worldsheet indices (here $\tau$,
$\sigma$). The worldsheet normals $n^i_{(\alpha)}$ satisfy
the relations
\begin{equation}
g_{ij} n^i_{(\alpha)}n^{j}_{(\beta)} = \delta_{\alpha\beta}
 ,\hspace{0.2in} g_{ij}n^i_{(\alpha)} e^j_{a}=0 \ ,
\end{equation}
where $\alpha=1..,N-2$ and $N$ is the dimension of the
background spacetime. The last condition holds for all $\alpha$
and $a$. Extrinsic curvature tensor components $K_{ab}^{(\alpha)}$ along
each normal $n^i_{(\alpha)}$ of the embedded worldsheet are
\begin{equation}
K_{ab}^{(\alpha)} = -g_{ij}( e^k_{a}\nabla_k e^i_b) n^{j(\alpha)}
\ .
\end{equation}
The equations of motion lead to the
condition, $K^{(\alpha)} = \gamma^{ab} K_{ab}^{(\alpha)} =0$
for an extremal worldsheet.
Thus extremal surfaces are those for which the trace of the extrinsic curvature tensor
along each normal is zero. Normal
deformations are denoted as
$\phi^{(\alpha)}$ along each normal. Therefore, the deformations constitute a
set of scalar fields. More explicitly, the deformation of each coordinate is
\begin{equation}
\delta x^i = \phi^{(\alpha)} n^i_{(\alpha)} \ ,
\end{equation}
which is the perturbation of the worldsheet (i.e. $x^i
\rightarrow x^i + \delta x^i$). For a worldsheet
with $\gamma^{ab}K_{ab}^{(\alpha)}=0$ satisfying the equations of
motion and the Virasoro constraints, the scalars
$\phi^i_{(\alpha)}$ satisfy the Jacobi
equations given as
\begin{equation}
\frac{1}{\Omega^2} \left ( - \frac{\partial^2}{\partial \tau^2} +
\frac{\partial^2}{\partial \sigma^2} \right ) \phi^{(\alpha)} +
\left (M^2\right )^{(\alpha)}_{(\beta)} \phi^{(\beta)} = 0 \ ,
\end{equation}
where
\begin{equation}
\left (M^2\right )^{(\alpha)}_{(\beta)} = K_{ab}^{(\alpha)}K^{ab}_
{(\beta)} +R_{ijkl} e^j_a e^{l\,\,a} n^{i\,\,(\alpha)}
n^k_{(\beta)} \ ,
\end{equation}
and $\Omega^2(\tau,\sigma)$ is the conformal factor of the
conformally flat form of the worldsheet line element. The Jacobi equations are
obtianed by constructing the second variation of the worldsheet action. 
Thus, solving these equations
for the perturbation scalars one can analyse the stability properties
of the extremal worldsheet. In other worlds, a stable worldsheet will
correspond to an oscillatory character for the $\delta x^i$ defined above. 

\noindent  Note further that the Jacobi
equations are like a family of coupled, variable `mass'  wave
equations for the scalars $\phi^{(\alpha)}$. They are 
a generalisation of the familiar geodesic deviation equation for 
geodesic curves in a Riemannian geometry. Usually, these equations 
are quite complicated and not easily solvable, even for the
simplest cases. It turns out that for the string configurations under consideration here,
we do find analytical solutions.

\noindent In a more general context, the
worldsheet covariant derivative
does have a term arising 
from the extrinsic twist potential (normal fundamental form)
which is given as:
$\omega_a^{\,\,\alpha\beta}=
g_{ij}\, \left ( e^k_a\nabla_k \, n^{i\,\alpha}\right )\,n^{j\,\beta}$.
For codimension one surfaces, i.e. hypersurfaces (as is the case here), the $\omega_a^{\alpha\beta}$
are all identically zero.

\subsection{The case of spiky strings in $AdS_3$}

Let us now turn to the string configurations mentioned 
earlier: i.e. the $AdS_3$ spiky strings. To proceed, we 
first write down the normal, induced metric and
extrinsic curvature for the world sheet configurations in the
Jevicki-Jin gauge stated earlier. 

\noindent The normal to the worldsheet is given as:
\begin{equation}
n^i = \left (\frac{\omega \rho' \tanh \rho}{\cosh^2\rho-\omega^2\sinh^2\rho}, -
\frac{\sinh 2\rho_0}{\sinh 2 \rho}, \frac{\rho' \coth \rho}{\cosh^2\rho-\omega^2\sinh^2\rho}\right )\, \, .
\end{equation}
\noindent Using the expression of $\rho(\sigma)$ 
one can write the normal in the
following form:
\begin{equation}
n^i = (n^0, ~n^1, ~n^2) \, ,
\end{equation}
where,
\begin{equation}
n^0 = \frac{\omega~ \sqrt{\delta~ \gamma~ \sn^2 (u,k)- 2\delta~ (\delta+1) + \frac{\eta~ \delta}{\sn^2(u,k)}}}{\alpha - \gamma~ \sn^2(u,k)}\, \, ,
\end{equation}

\begin{equation}
n^1 = \frac{-\sinh 2\rho_0}{\sqrt{\gamma^2~ \sn^4(u,k) - 2 \gamma~ (\delta+1)~ \sn^2(u,k) + \alpha~ \delta}} \, \, ,
\end{equation}

\begin{equation}
n^2 = \frac{\sqrt{\delta~ \gamma~ \sn^2 (u,k)- 2\delta ~(\delta+1) + \frac{\eta~ \delta}{\sn^2(u,k)}}}{\delta - \gamma~ \sn^2(u,k)} \, \, ,
\end{equation}
and $\alpha, ~\delta , ~\gamma ~\& ~\eta$ are defined as follows
$$\alpha = 1 + \cosh 2\rho_1, ~\delta = \cosh 2\rho_1 -1,~ \gamma = \cosh 2\rho_1 - \cosh 2\rho_0, ~\&  ~\eta = \cosh 2\rho_1 + \cosh 2\rho_0 \ .$$

The extrinsic curvature tensor turns out to be
\begin{equation}
K_{ab} = \begin{pmatrix} \frac{1-\omega^2}{2} \sinh 2\rho_0 & -\omega \cr -\omega & \frac{1-\omega^2}{2} \sinh 2\rho_0 \end{pmatrix} .
\end{equation}
The induced metric is given as:
\begin{equation}
ds^2 = \left (\cosh^2\rho-\omega^2\sinh^2\rho\right )\left (-d\tau^2+d\sigma^2\right).
\end{equation}
Hence the quantity $K_{ab}K^{ab}$ is given as:
\begin{equation}
K_{ab} K^{ab} = 2\frac{\left (1-\omega^2\right)^2  \sinh^2\rho_0 \cosh^2\rho_0-\omega^2}{\left (\cosh^2 \rho-\omega^2\sinh^2\rho\right )^2}= \frac{P}{\Omega^4} \, \, ,
\end{equation}
%Note that $Q$ is never equal to zero.
where $P =2\left (1-\omega^2\right)^2  \sinh^2\rho_0 \cosh^2\rho_0-\omega^2 $ and $\Omega^2 = \left (\cosh^2 \rho-\omega^2\sinh^2\rho\right )$.
Similarly one can write down the contribution from the
Riemann tensor term in the perturbation equation as follows
\begin{equation} 
R_{ijkl} e^j_a e^{l\,\,a} n^{i\,\,(\alpha)}n^k_{(\beta)}= -2 ~\delta^{\alpha}_{\beta} \ .
\end{equation}
 Using all of the above-stated quantities which appear
in the perturbation equation and after some lengthy algebra, 
one arrives at the
following equation for the perturbation scalar $\phi$
\begin{equation}
\left ( -\frac{\partial^2}{\partial \tau^2} +
\frac{\partial^2}{\partial \sigma^2} \right )\phi+ 
 \left [ -2 \Omega^2 + \frac{P}{\Omega^2}
\right  ] \phi =0 \, \, .
\end{equation}
Let us now use the functional form of $\rho(\sigma)$ and the ansatz
\begin{equation}
\phi(\tau,\sigma) = \epsilon
e^{i\beta \tau} R(\sigma) \, \, ,
\end{equation}
where $\beta$ will be the eigenvalue and $\epsilon$, a constant which we may relate to the
amplitude of the perturbation. It must be emphasized that $\epsilon$
has to be small in value (i.e. $\epsilon<<1$) in order to ensure that the
deformation is genuinely a perturbation. With the above ansatz for $\phi(\tau,\sigma)$ one can reduce the perturbation equation to
an equation for $R(\sigma)$ given as:
\begin{equation}
\frac{d^2 R}{du^2} + \left [\beta_1^2-2 \left (k^2 \sn^2(u,k) + \frac{1}{\sn^2(u,k)}\right )\right ] R =0 \, \, ,
\label{ep1}
\end{equation}
or
\begin{equation}
\frac{d^2 R}{du^2} + \left [\beta_1^2-V(u)\right ] R =0 \, \, ,
\label{ep10}
\end{equation}
where $\beta_1^2= \frac{\beta^2}{\zeta^2}$, $u=\zeta \sigma$ (see Eqn. (\ref{z1})
for $\zeta$) and 
\begin{equation}
V(u) = 2 \left (k^2 \sn^2(u,k) + \frac{1}{\sn^2(u,k)}\right ).
\end{equation}
Eqn. (\ref{ep1}) is a special case of the well-known 
Darboux-Treibich-Verdier (DTV) equation \cite{sparre:1883}, \cite{matveev} of mathematical
physics. In general, the DTV equation has the following form:
\begin{equation}
\frac{d^2 R}{dx^2} + (h - V(x)) R = 0 \, \, ,
\label{ep2}
\end{equation}
 with $V(x)$ given as:
\begin{equation}
\begin{split}
V(x)& = \nu~(\nu+1)k^2~\sn^2(x,k)~+~\mu~(\mu+1)k^2~\frac{\cn^2 (x,k)}{\dn^2 (x,k)}~ 
+\eta~(\eta+1) \frac{\dn^2 (x,k)}{\cn^2(x,k)} \\
& +\xi~(\xi+1)\frac{1}{\sn^2 (x,k)} \ .
\end{split}
\label{ep3}
\end{equation}

\noindent The potential $V(x)$ in the DTV equation is a periodic potential which includes the well-studied Lam\'e potential as a special case. Our equation is another special case of Eqn. (\ref{ep3}) with $\nu = \xi = 1$ and 
$\mu = \eta = 0$ with period $2K$ where $K$ is defined as follows:
\begin{equation}
K = \int_{0}^{\frac{\pi}{2}} \frac{d \psi}{1 - k^2 \sin^2{\psi}} \, \ .
\end{equation}
In Figures 3 and 4, we show the nature of our potential for the cases with ten and three spikes, respectively. Notice these are well--like
potentials with a flat bottom (much like a generalisation of a
harmonic oscillator potential with added anharmonicities). It is
therefore obvious that there will be bound state solutions which will
correspond to the normal mode fluctuations of the spiky string.
\begin{center}
\begin{figure}[h]\label{f5}
 \begin{minipage}{14pc}
 \includegraphics[width=17pc]{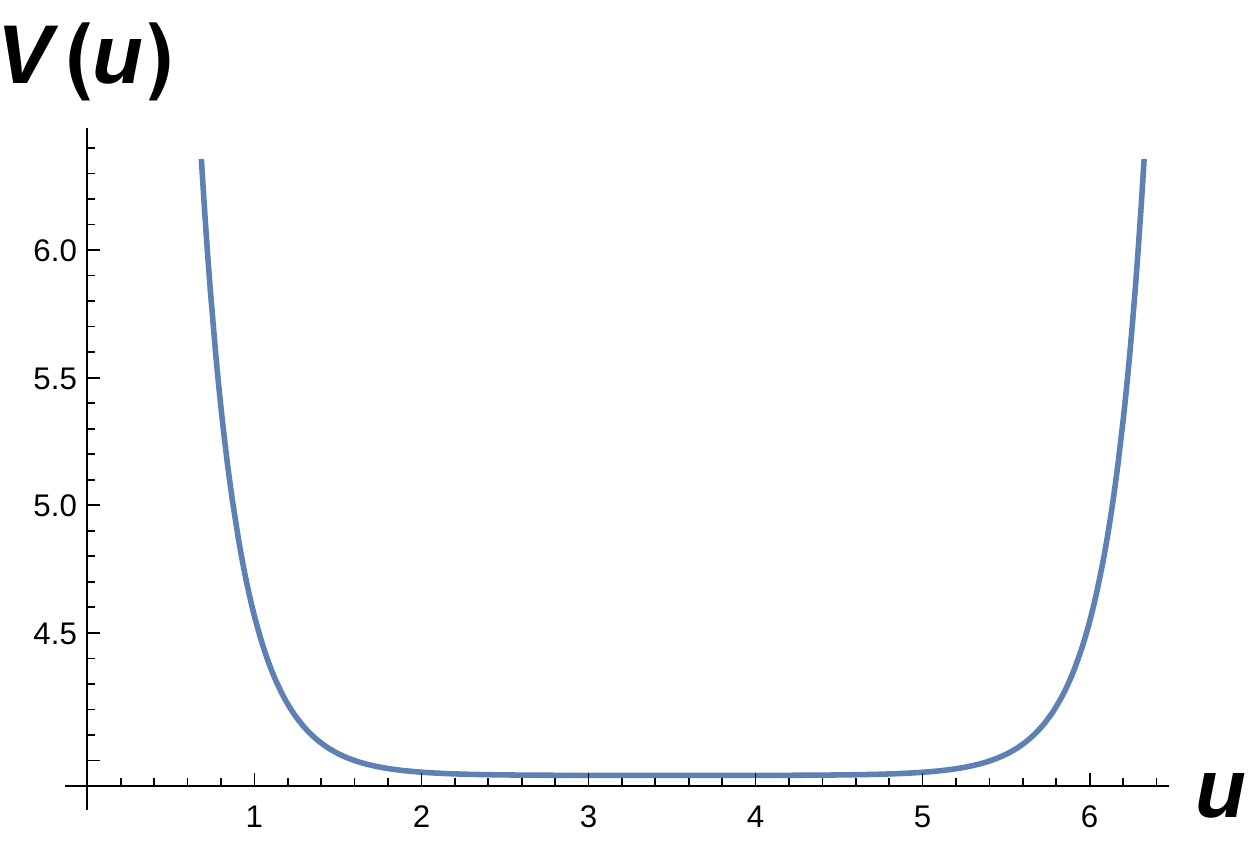}
 \caption{Plot of the potential $V(u)$ for one period range.}
\end{minipage}
\hspace{1.0in}
\begin{minipage}{14pc}\label{f6}
 \includegraphics[width=17pc]{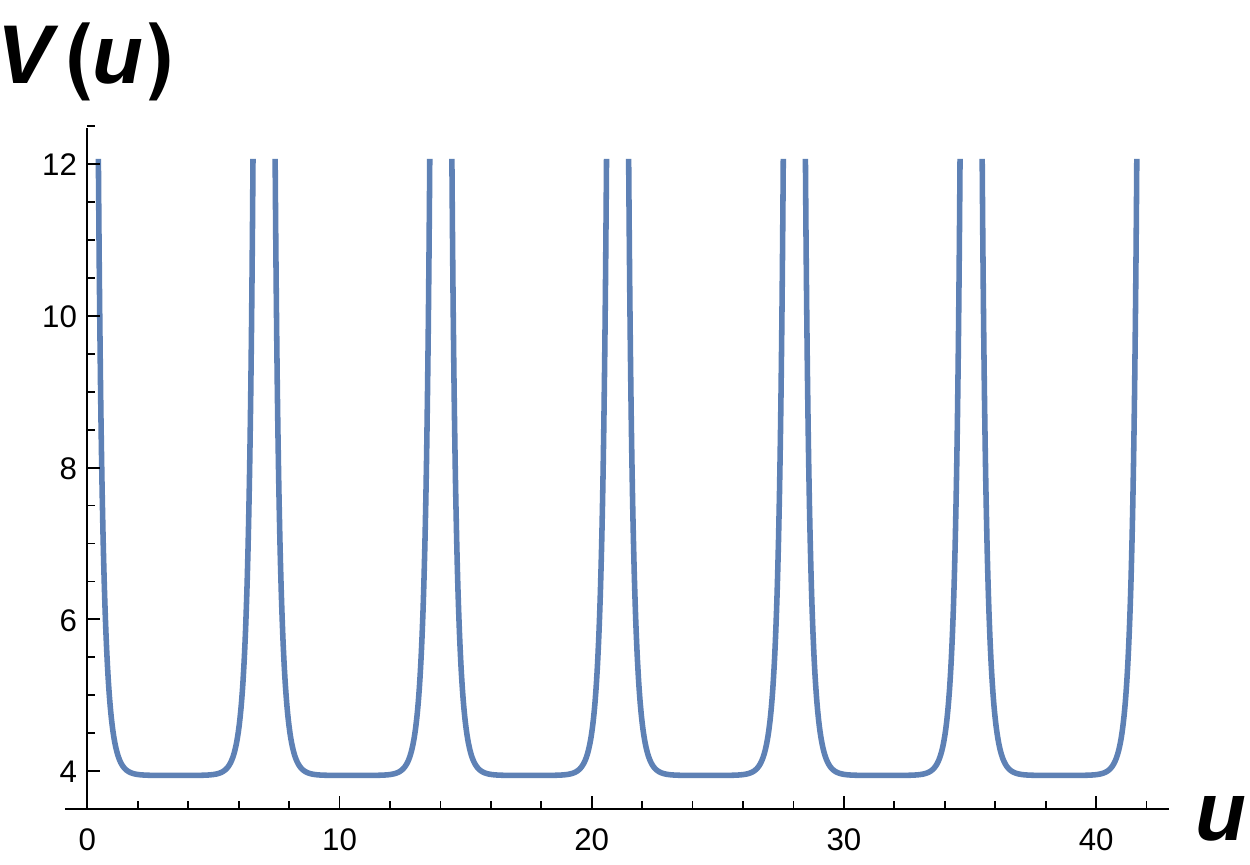}
 \caption{Plot of the potential $V(u)$ for more than one period range.}
\end{minipage}
\end{figure}
\end{center}
We will now move on to constructing solutions of the DTV equation
(for our special case) and thereafter analyse the
perturbations. This will lead us towards figuring out whether the spiky string configurations in $AdS_3$ 
are indeed stable or not. 

\subsection{Solving the Jacobi equation and the stability issue}

\noindent Various forms of the linearly independent general solutions 
to the DTV equation exist in the literature.
The earliest reference dates back to 1883 in a couple of articles by the mathematician Comte de Sparre \cite{sparre:1883, spr}, who seems to have
studied the DTV equation and its special cases in great detail.
More recently, in work by Matveev and Smirnov \cite{matveev} (who have rediscovered the work of Sparre), the general solutions have been written down
using products of Jacobi and other Theta  ($\Theta$) functions. We will however look at a 
series solution obtained recently in the article by Chiang et al \cite{ching} which turns out to be quite useful in obtaining the eigenfunctions and eigenvalues for our special case.  This, as we
shall see, further helps in understanding the stability
issue of spiky strings in $AdS_3$.

\noindent  A series solution of the general DTV equation ( $\xi \neq -\frac{3}{2}, -\frac{5}{2}$) as pointed out in \cite{ching} is given as: 
\begin{equation}
\sn^{\xi+1}(u,k)\cn^{\eta+1}(u,k)\dn^{\mu+1}(u,k)\sum_{m=0}^\infty C_m\sn^{2m}(u,k) \ .
\label{s1}
 \end{equation}
This expansion is known as a local Darboux solution and denoted as $Dl(\xi,\eta,\mu,\nu;h;u,k)$ where the $C_m$ satisfies the following recursion relation:
\begin{equation}
\begin{split}
&(2m+2)(2m+2\xi+3)C_{m+1}\\
&+\{h-[2m+\eta+\xi+2]^2-k^2[2m+\mu+\xi+2]^2+(k^2+1)(\xi+1)^2\}C_{m}\\
& +k^2(2m+\xi+\eta+\mu+\nu+2)(2m+\xi+\eta+\mu-\nu+1)C_{m-1}=0  \ .
\end{split}
\label{r1}
\end{equation}
For simplicity, one may define
$$M_m(\xi):=(2m+2)(2m+2\xi+3);$$
$$L_m(\xi,\,\eta,\mu;h;k):=h-[2m+\eta+\xi+2]^2-k^2[2m+\mu+\xi+2]^2+(k^2+1)(\xi+1)^2;$$
$$K_m(\xi,\eta,\mu,\nu;k):=k^2(2m+\xi+\eta+\mu+\nu+2)(2m+\xi+\eta+\mu-\nu+1)$$
%\begin{equation}
so that (\ref{r1}) takes the form
\begin{equation}
M_mC_{m+1}+L_mC_m+K_mC_{m-1}=0 \ .
\end{equation}

\noindent We state below the content of two theorems regarding the convergence of  $Dl(\xi,\eta,\mu,\nu;h;u,k)$ \cite{ching}.

\noindent The first of these theorems as quoted in \cite{ching}
states the following. If there exists a positive integer $q$ such that either
\begin{equation}
\xi+\eta+\mu+\nu=-2q-4\quad\mbox{ or }\quad \xi+\eta+\mu-\nu=-2q-3,
\end{equation}
holds, then there exist $q+1$ values $h_0,\, \cdots,\, h_q$ of $h$ such that the 
series $Dl(\xi,\eta,\mu,\nu;h_j;u,k)$ ($j=0,1,\cdots,q$) terminates and yields a
polynomial.

\noindent In the second theorem, the following result is stated.
Suppose that $Dl(\xi,\eta,\mu,\nu;h;u,k)$ is non-terminating, i.e., not a Darboux polynomial. Then it converges on the domain $\{|\sn u|<
\max(1,|k|^{-1})\}$ ($|k|\neq1$) if 
\begin{equation}
 g(\xi,\eta,\mu,\nu;h;k):=L_0/M_0-\frac{K_1/M_1}{L_1/M_1-}\frac{K_2/M_2}{L_2/M_2-}\cdots=0
\label{cf1}
\end{equation}
holds. Otherwise, it converges only on the domain $\{|\sn u|<\min(1,|k|^{-1})\}$.

\noindent Both these theorems are crucial in establishing the solution and
analysing its nature. In our case however, we do not end up with a polynomial.
However, the structure of the infinite series term is such that there is no
question of any instability in the final form of the deformations $\delta x^i$, as
we shall show in the forthcoming discussion.

\noindent Eqn. (\ref{ep2}) is a Schr\"{o}dinger-like equation with a periodic potential with period $2 K$. So the $\sigma$ dependent perturbation parameter $R(\sigma)$
will have to satisfy the Bloch condition which gives $ 2 K = 2 \pi \zeta$. Putting values we get
\begin{equation}
K (k^2) = \pi \sqrt {\frac{\cosh 2\rho_1 + \cosh 2\rho_0}{\cosh 2\rho_1 -1}} \ .
\label{cp1}
\end{equation}

\noindent Since we are dealing with a closed string with spikes the closedness condition must also be satisfied. If the angle between the spike and the valley is $\Delta \psi$, and the 
no of spikes is $n$, then the closedness condition is satisfied through the following equation
\begin{equation}
\Delta \psi (\rho_1 , \rho_0) = \frac{2 \pi}{2 n} \, \, ,
\label{cp2}
\end{equation}
where $\Delta \psi$ has the following form
\begin{equation}
%\begin{split}
 \Delta \psi (\rho_1, \rho_0) =  \frac{\sinh2\!\rho_0}{\sqrt{2}\sinh\rho_1}\frac{ \left\{\Pi(\frac{\pi}{2},\frac{\cosh 2\rho_1-\cosh 2\rho_0}{\cosh 2\rho_1-1},p)  
-\Pi(\frac{\pi}{2},\frac{\cosh 2\rho_1-\cosh 2\rho_0}{\cosh 2\rho_1+1},p)\right\}}{\sqrt{\cosh 2\rho_1+\cosh 2\rho_0}} \, \, ,
 %\end{split}
%\hspace{0.5in}
\end{equation} 
where $p$ is defined as
\begin{equation}
p = \sqrt{\frac{\cosh 2 \rho_1 - \cosh 2 \rho_0}{\cosh 2\rho_1 + \cosh 2 \rho_0}} \ .
\end{equation}

\noindent Thus, Eqns. (\ref{cp1}) and (\ref{cp2}) should be satisfied simultaneously by $\rho_1 ~\& ~\rho_0$ to respect the periodicity and closedness conditions, respectively. Thus, for a fixed number of 
spikes $n_0$ one can get the value of  $\rho_1 ~\& ~\rho_0$ from Eqns. (\ref{cp1}) and (\ref{cp2}). In one of our cases ($n = 10$ ), solving Eqns. (\ref{cp1}) and (\ref{cp2}) 
we get $\rho_1 = 3.18 ~\& ~\rho_0 = 1.068$. Using these values we get $k = 0.985264$. In another case ($n = 3$) we get $\rho_1 = 2.645$ ~\& $\rho_0 = 0.45$ and $k = 0.985653$.
Similarly, we can obtain the values for any other $n$ too.

\noindent In our problem, we have $\xi= \nu= 1$ \& $\eta= \mu = 0$. One can easily check the first of the two results quoted above does not hold (no termination of the series). Hence we need to look at the next result. 
From the expressions of $L_m,~ M_m, ~\& ~K_m$ one can see that if we put the value of $\xi, ~\nu, ~\eta ~\& ~\mu$ then the value of $L_m ~\&~ M_m$ will be fixed 
if we put the value of $m$ but $K_m$ depends on the eigenvalue $h$. Thus the 
truncated infinite continued fraction (\ref{cf1}) effectively gives us a polynomial equation for $h$ and to get 
a convergent  $Dl(\xi,\eta,\mu,\nu;h;u,k)$, the only allowed $h$ are the roots of the polynomial. We therefore compute the continued fraction by truncating it at some order 
and thereby obtain the roots. However, to satisfy the convergence criteria, 
truncation should be made at a sufficiently higher order, as we will demonstrate
explicitly.

\begin{center}
\begin{figure}[h]
 \begin{minipage}{14pc}
 \includegraphics[width=17pc]{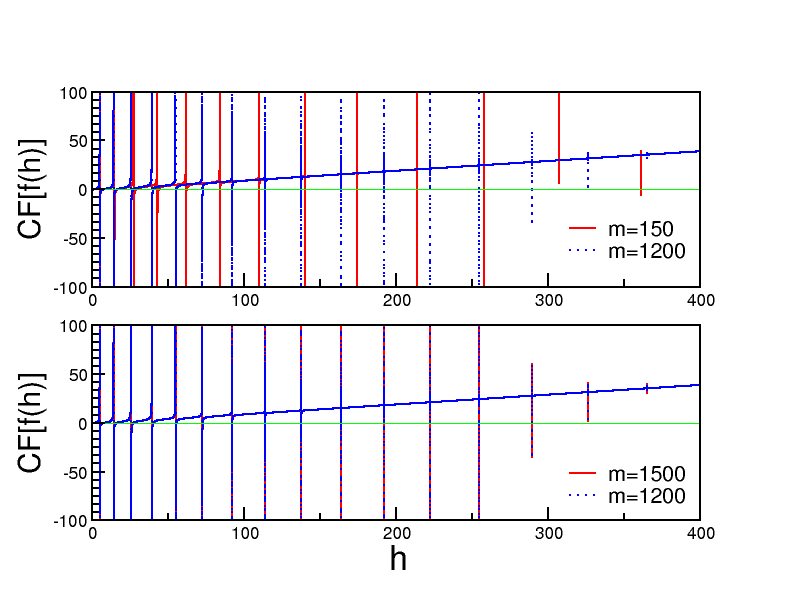}
 \caption{Plot of the continued fraction $CF[f(h)]$ for $n=10$ spikes. In the upper figure the red curve is for $m = 150$ and the blue is for $m = 1200$. In the lower figure the red curve is for $m = 1500$ and the blue one is for $m = 1200$.}
\label{f1}
\end{minipage}
\hspace{1.2in}
 \begin{minipage}{14pc}
 \includegraphics[width=17pc]{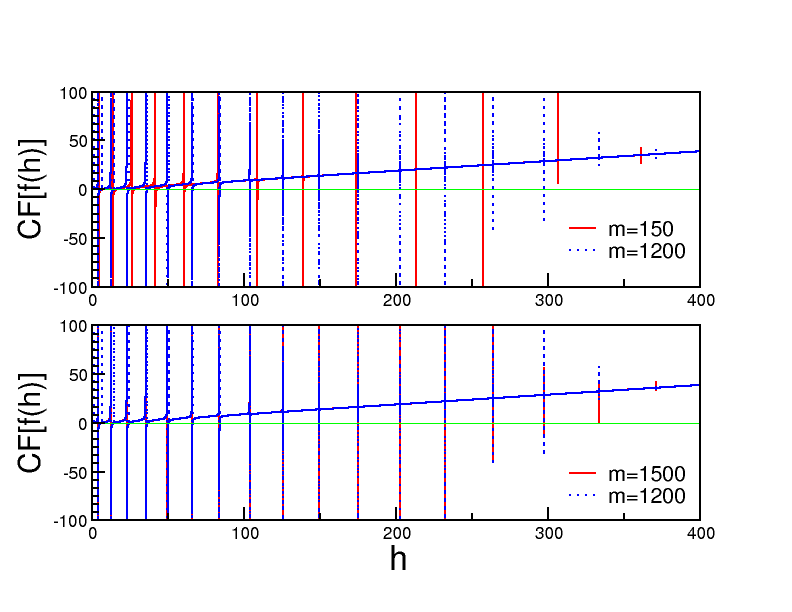}
 \caption{Plot of the continued fraction $CF[f(h)]$ for $n=3$ spikes. In the upper figure the red curve is for $m = 150$ and the blue is for $m = 1200$. In the lower figure, the red curve is for $m = 1500$ and the blue one is for $m = 1200$.}
\label{f2}
\end{minipage}
 \end{figure}
\end{center}

\begin{center}
\begin{figure}[h] 
 \begin{minipage}{14pc}
 \includegraphics[width=17pc]{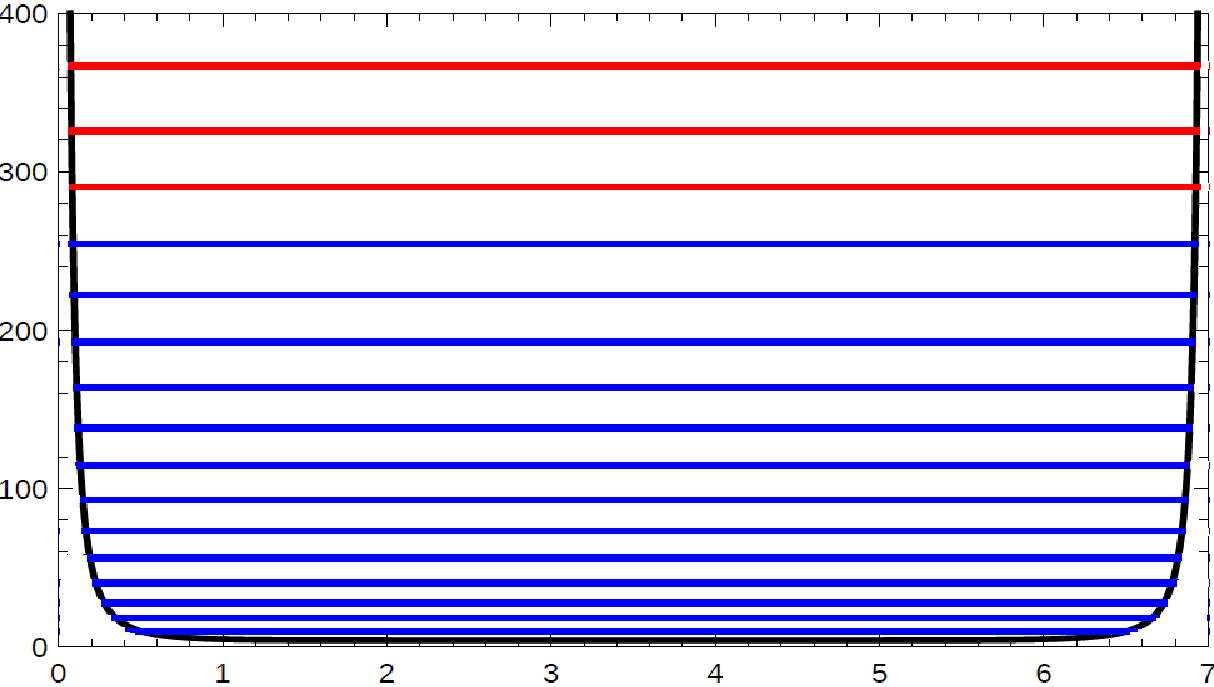}
 \put(0,5){$\bf u$}
 \put(-200,120){$\bf {V(u)}$}
 \caption{Plot of  different roots in the potential $V(u)$ for $n=10$ spikes. The top three levels which we have used in our calculations, starting from top have $h=366.6$ (red), $h = 326$ (red)  \& $h = 290.2$ (red) respectively.}
\label{v8}
\end{minipage}
\hspace{1.0	in}
\begin{minipage}{14pc}
 \includegraphics[width=17pc]{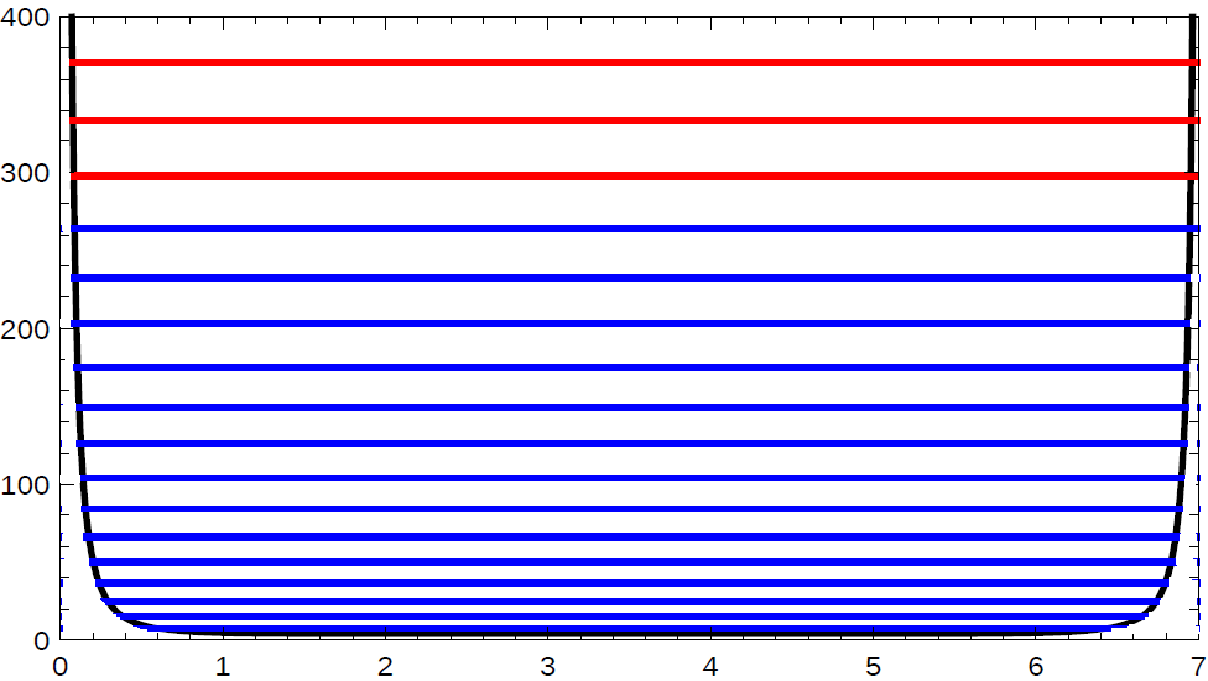}
 \put(0,5){$\bf u$}
 \put(-200,120){$\bf {V(u)}$}
 \caption{Plot of  different roots in the potential $V(u)$ for $n = 3$ spikes. The top three levels which we have used in our calculations, starting from top have $h=370.74$ (red), $h = 333.23$ (red)  \& $h = 297.87$ (red) respectively.}
\label{v9}
\end{minipage}
\end{figure}
\end{center}

\noindent We solve the continued fraction numerically by using double precision. Figure (\ref{f1}) shows the plot of the continued fraction for two different $m$, $m = 150$ (shown by a red curve) 
and $m = 1200$ (shown by a blue curve). The $h$ values at which the graph intersects the $h$ axis, provides us with the roots. In Figure (\ref{f1}) one can see that there is some mismatch between the roots of the $m=150$ order truncation and $m=1200$ order truncation of the
continued fraction.  One can observe  this by noting that the red curve and the blue curve cut the axis at different points. A convergence to the roots of the infinite continued fraction is reached when there is no mismatch between two truncations at different orders. Or, in other words we can say that if one considers two separate  polynomials of $h$ of different orders and finds that the roots for the two different polynomials are very close in value to each other, then there is
convergence. However, in the above example we can see that there is a mismatch and the values found with a truncation at a lower order (i.e at $m=150$) cannot be reliable . It is therefore certain from the above 
figure that to achieve convergence one has to truncate the infinite continued fraction at sufficiently higher order. Figure (\ref{f2}) shows the truncated continued fraction for $m=1200$ and $m=1500$.  It is
clear that there is almost no such mismatch as observed in Figure (\ref{f1}). The blue curves ($m=1200$) and the red curves ($m=1500$) overlap completely at least up to some finite domain of $h$. If one truncates   at further higher order, the convergence  will be even better. 
%One can easily see that roots do not converge and the convergence is poorer for 
%higher roots. On the contrary, Figure (\ref{f2}) shows the continued fraction plot for $m = 1500$ (red line)
%and $m = 1200$ (green line). Here one can see that the roots are converging (no distinct red
%and green curves here)
%because we have truncated at a sufficiently high order.

\begin{center}
\begin{figure}[h]
\begin{subfigure}[Plot for $n=10$ with $h = 366.6$ (red), $h = 326$ (blue), \& $h = 290.2$ (green)]
 {\includegraphics[width=17pc]{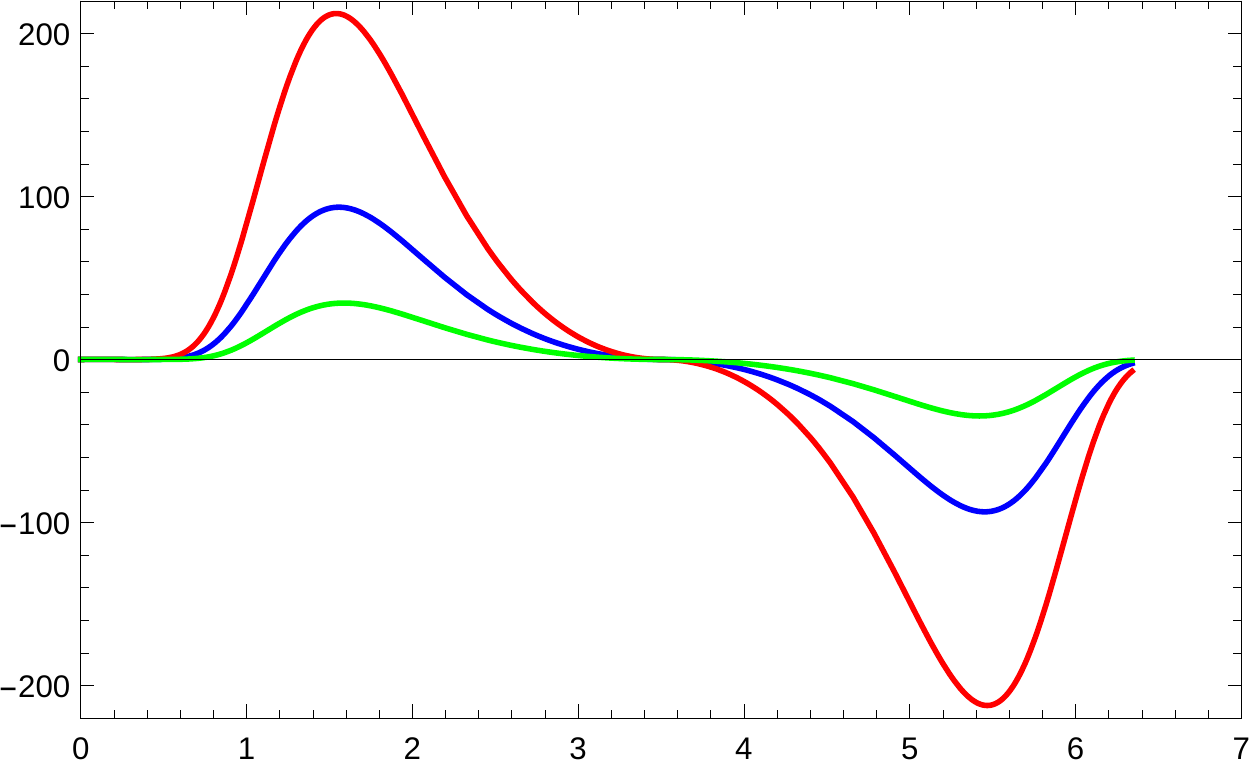}}
 \put(0,60){$\bf u$}
 \put(-190,130){$\bf {\delta t}$}
\label{f3}
\end{subfigure}
\hspace{0.5in}
\begin{subfigure}[Plot for $n=3$ with $h=370.74$ (red), $h =333.23$ (blue) , \& $h = 297.87$ (green)]
 {\includegraphics[width=17pc]{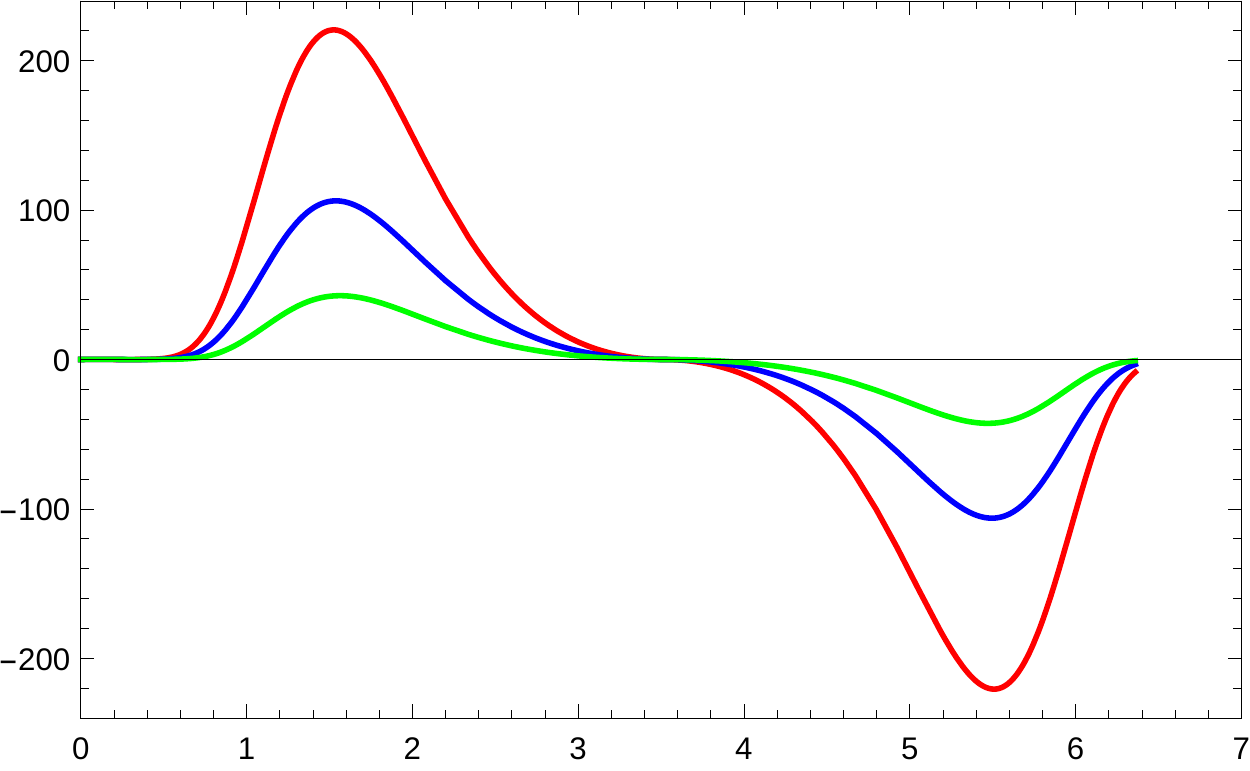}}
 \put(0,60){$\bf u$}
 \put(-190,130){$\bf {\delta t}$}
\label{f10}
\end{subfigure}
\caption{Plot of $\delta t$ for $n=10 ~\& ~3$ spikes for different $h$}
\label{l1}
\end{figure}
\end{center}
\begin{center}
\begin{figure}[h]
\begin{subfigure}[Plot for $n=10$ with $h=366.6$ (red), $h = 326$ (blue) , \& $h = 290.2$ (green)]
{\includegraphics[width=17pc]{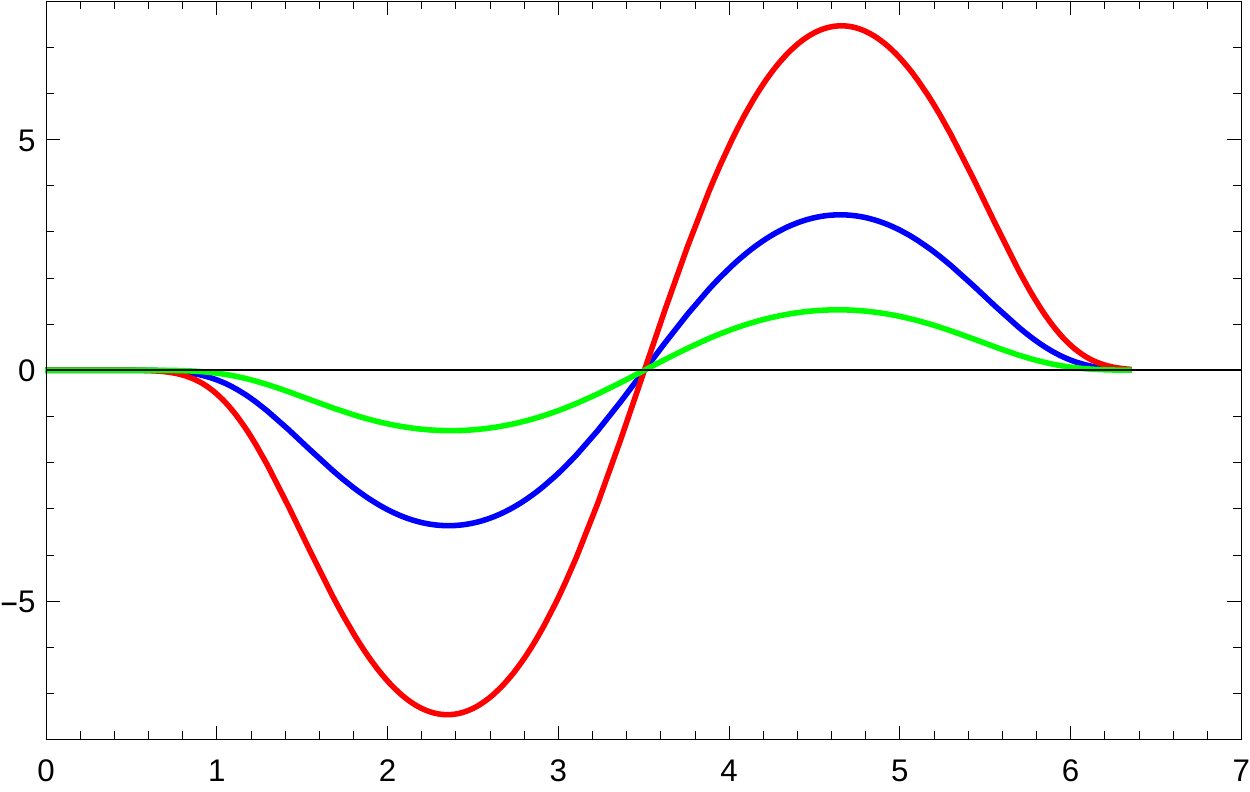}}
\put(0,60){$\bf u$}
 \put(-190,135){$\bf {\delta \rho}$}
\label{f4}
\end{subfigure}
\hspace{0.5in}
\begin{subfigure}[Plot for $n=3$ with $h=370.74$ (red), $h = 333.23$ (blue) , \& $h = 297.87$ (green)]
{\includegraphics[width=17pc]{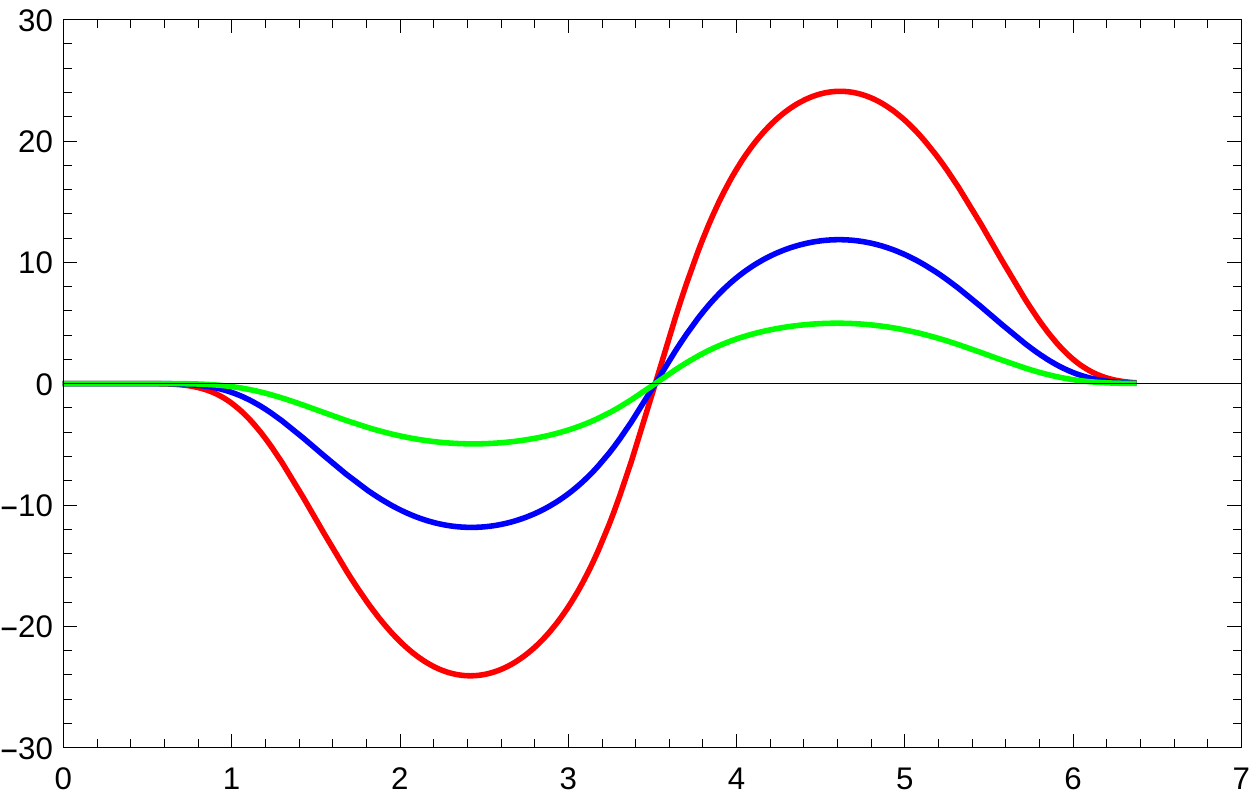}}
\put(0,60){$\bf u$}
 \put(-190,135){$\bf {\delta \rho}$}
\label{f11}
\end{subfigure}
\caption{Plot of $\delta \rho$ for $n = 10 ~\& ~3$ spikes for different $h$}
\label{l2}
\end{figure}
\end{center}
\begin{center}
\begin{figure}[h]
\begin{subfigure}[Plot for $n=10$ with $h=366.6$ (red), $h = 326$ (blue) , \& $h = 290.2$ (green)]
 {\includegraphics[width=17pc]{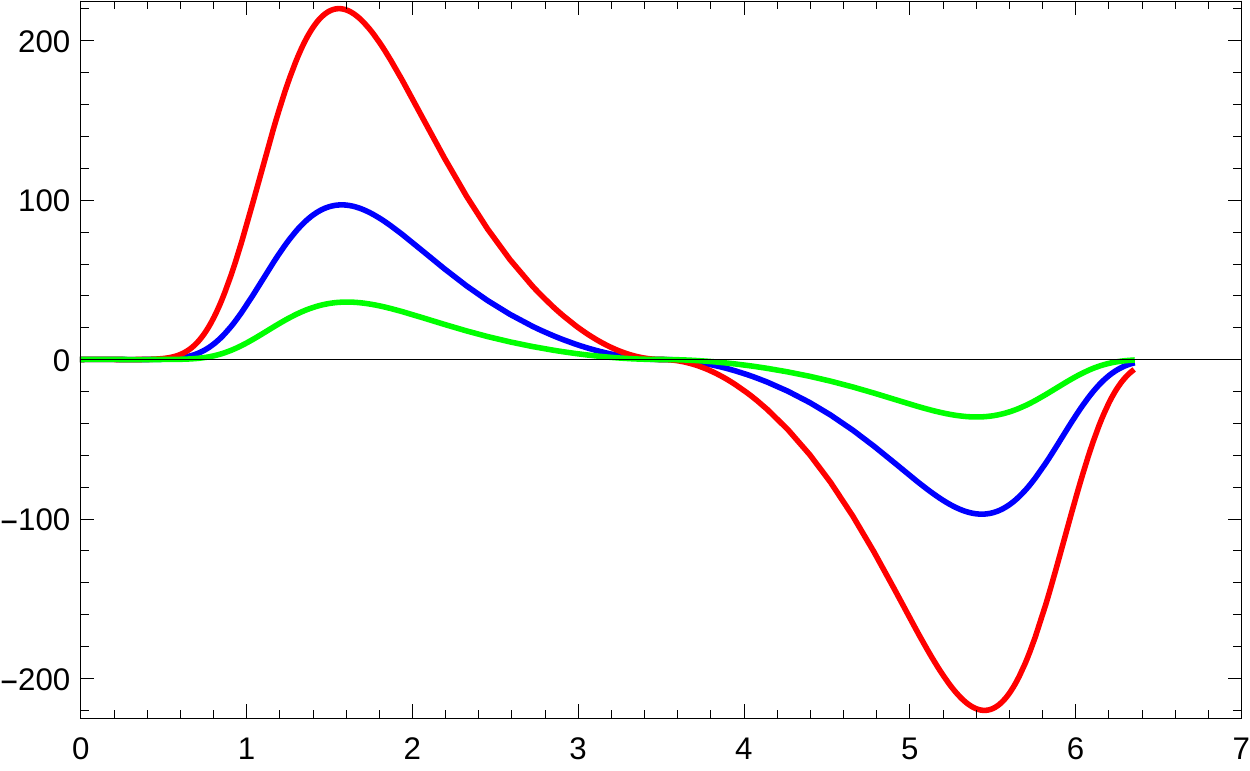}}
 \put(0,60){$\bf u$}
 \put(-190,130){$\bf{\delta} \bf {\theta}$}
 \label{f7}
\end{subfigure}
\hspace{0.5in}
\begin{subfigure}[Plot for $n=3$ with $h=370.74$ (red), $h = 333.23$ (blue) , \& $h = 297.87$ (green)]
 {\includegraphics[width=17pc]{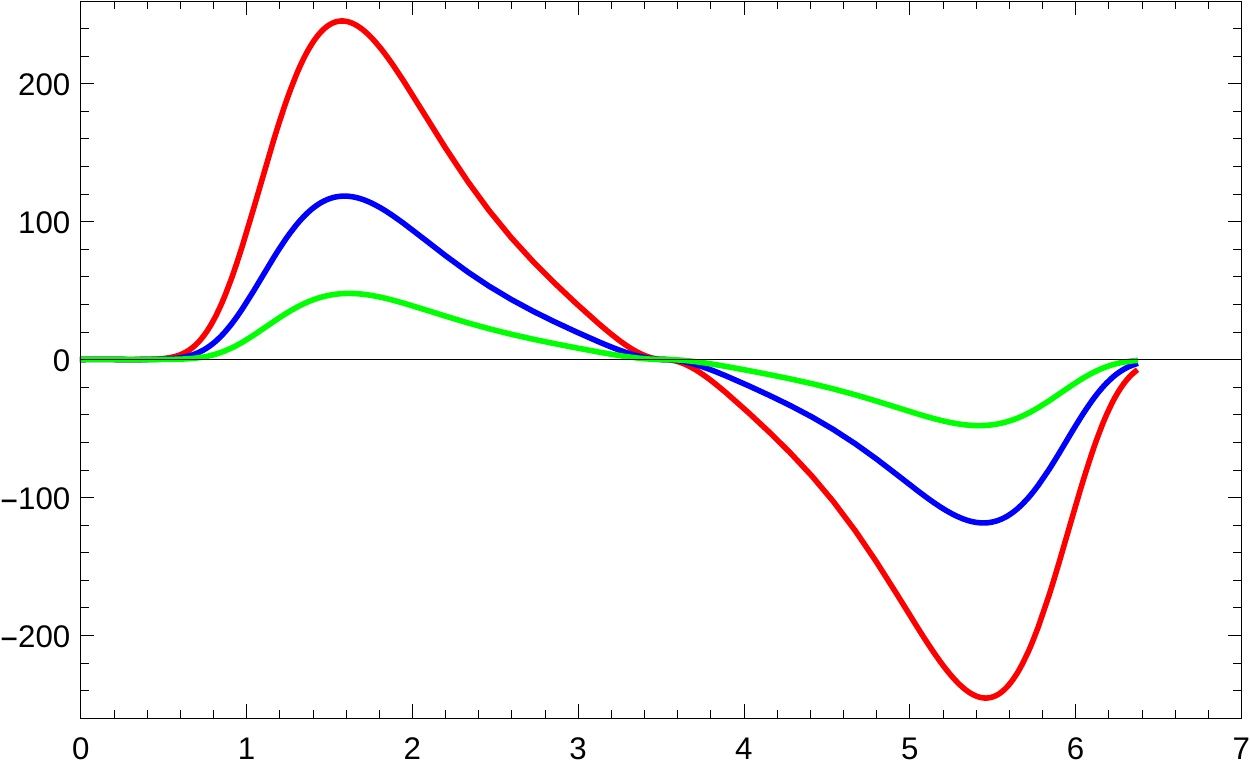}}
 \put(0,60){$\bf u$}
 \put(-190,130){$\bf{\delta} \bf {\theta}$}
\label{f12}
\end{subfigure}
\caption{Plot of $\delta \theta$ for $n = 10 ~\& ~3$ spikes for different $h$}
\label{l3}
\end{figure}
\end{center}

\begin{center}
\begin{figure}[h] 
 \begin{minipage}{14pc}
 \includegraphics[width=17pc]{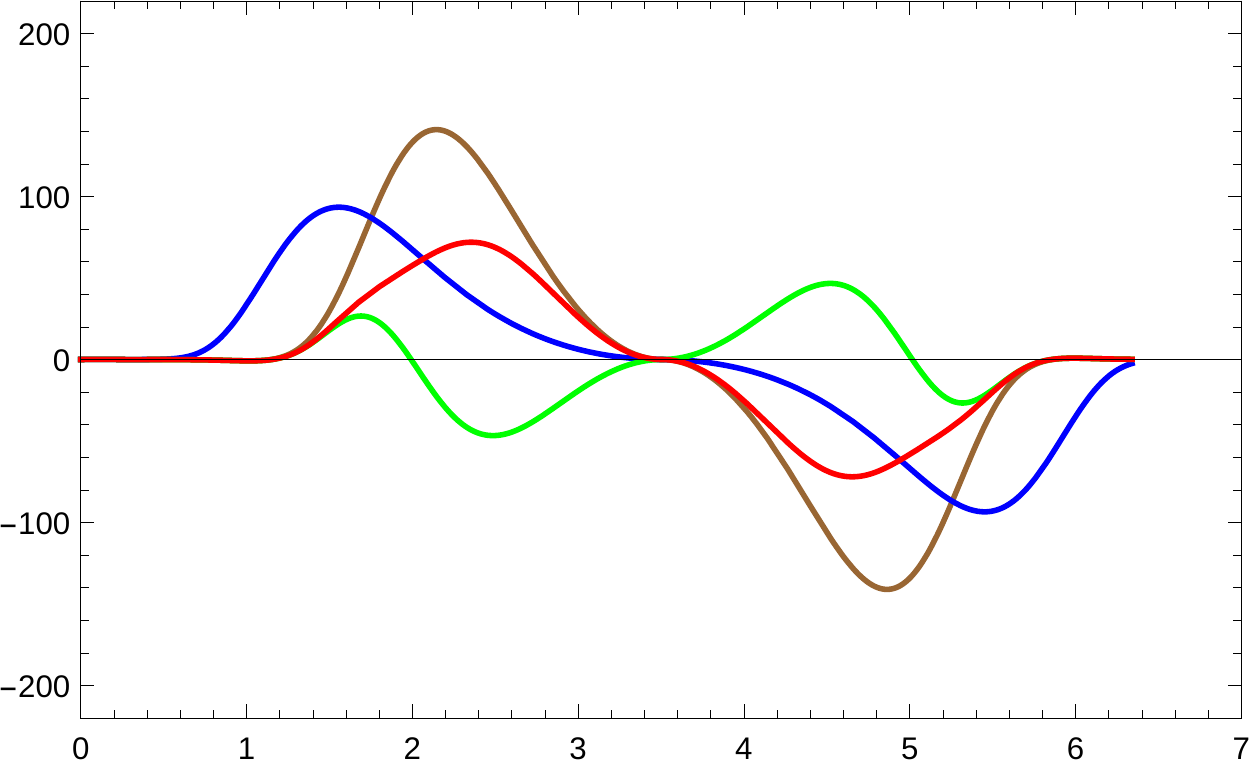}
 \put(0,65){$\bf u$}
 \put(-200,135){$\bf {\delta t}$}
 \caption{Plot of $\delta t$ for $h=326$ for four different number of terms i) $N=5$ (blue), ii) $N=21$ (brown), iii) $N=33$ (green) \& iv) $N=43$ (red)}
\label{c1}
\end{minipage}
\hspace{1.0in}
\begin{minipage}{14pc}
 \includegraphics[width=17pc]{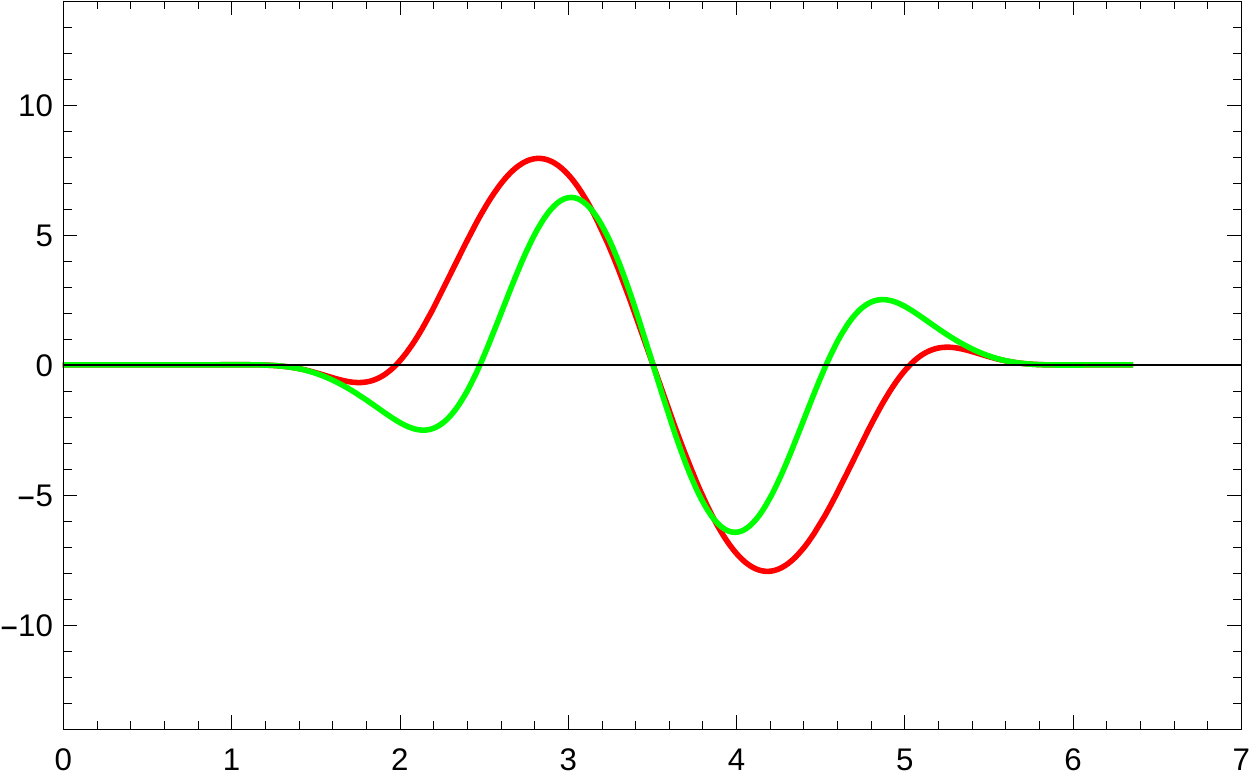}
 \put(0,65){$\bf u$}
 \put(-200,135){$\bf {\delta \rho}$}
 \caption{Plot of  $\delta \rho$ for two negative ranges, $N$ = 33 (red) \& $N$ = 73 (green).}
\label{c2}
\end{minipage}
\end{figure}
%\hspace{0.1in}
\begin{figure}[h]
\begin{minipage}{14pc}
 \includegraphics[width=17pc]{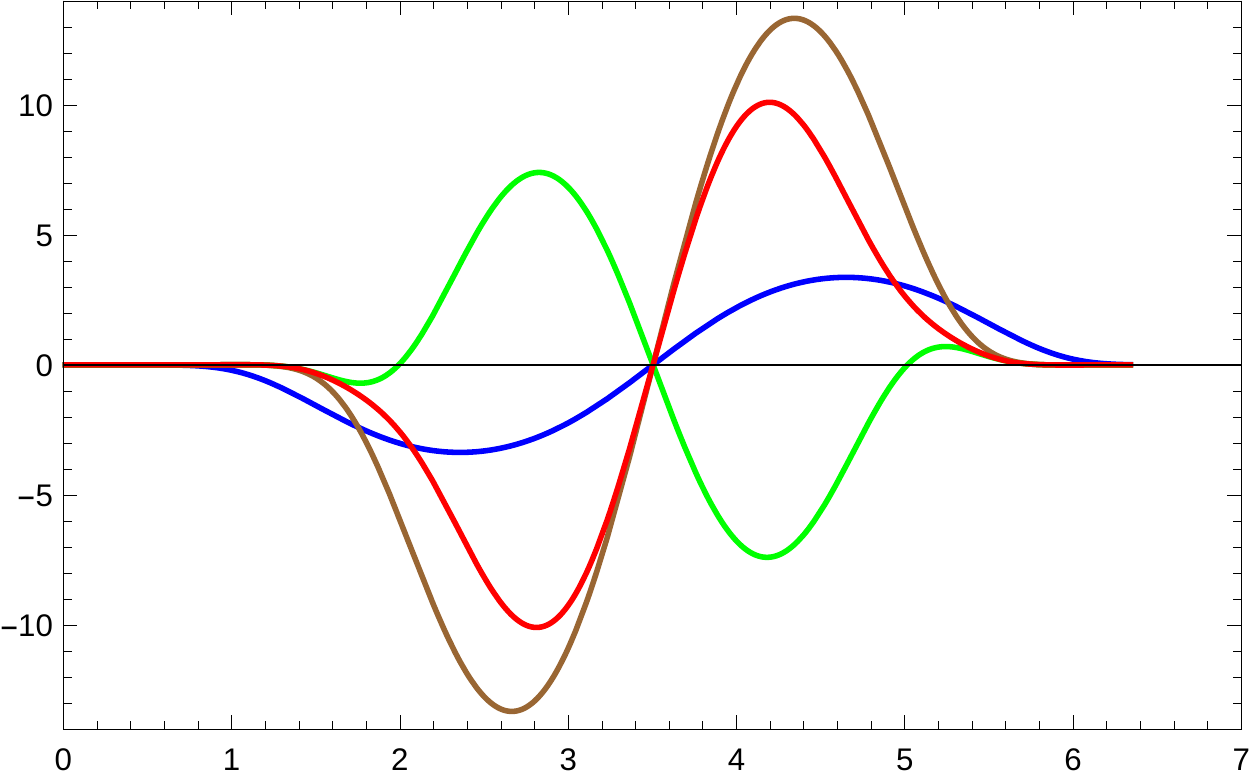}
 \put(0,65){$\bf u$}
 \put(-200,135){$\bf {\delta \rho}$}
 \caption{Plot of $\delta \rho$ for $h=326$ for four different number of terms i) $N=5$ (blue), ii) $N=21$ (brown), iii) $N=33$ (green) \& iv) $N=43$ (red)}
\label{c3}
\end{minipage}
\hspace{1.0in}
\begin{minipage}{14pc}
 \includegraphics[width=17pc]{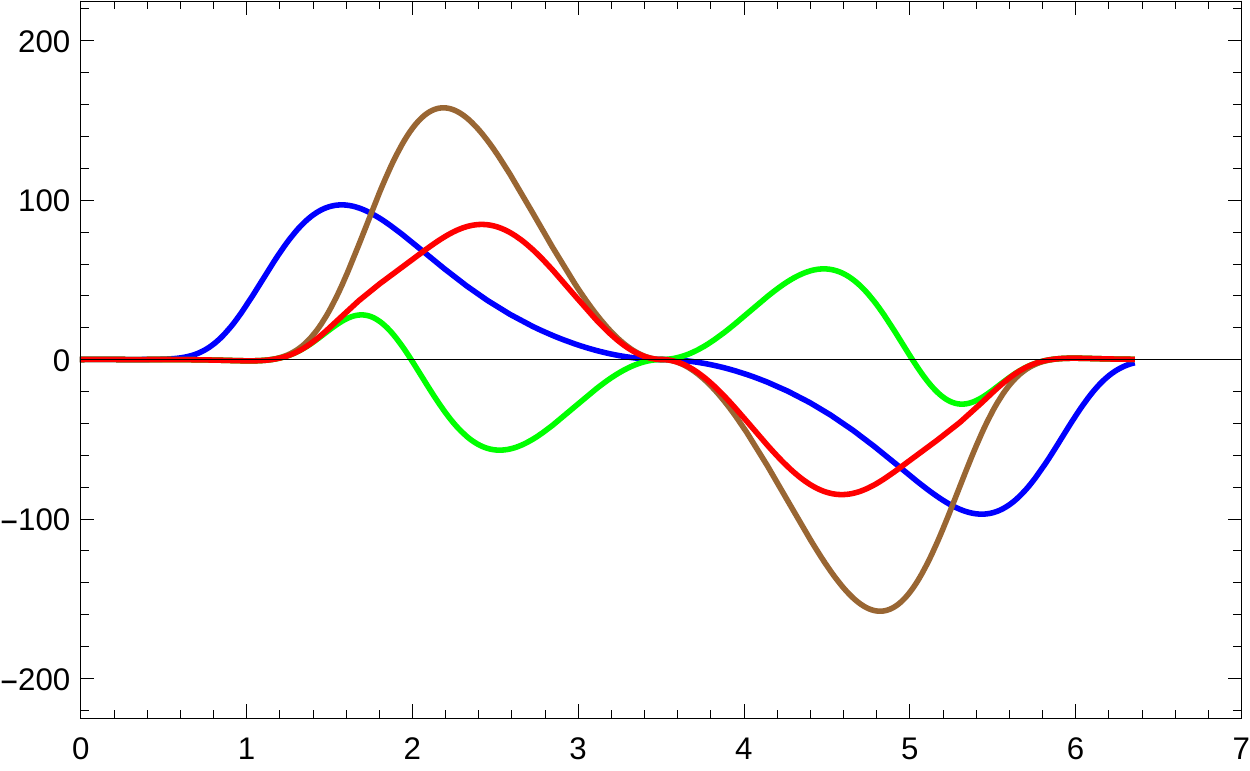}
 \put(0,65){$\bf u$}
 \put(-200,135){$\bf {\delta \theta}$}
 \caption{Plot of $\delta \theta$ for $h=326$ for four different number of terms i) $N=5$ (blue), ii) $N=21$ (brown), iii) $N=33$ (green) \& iv) $N=43$ (red)}
\label{c4}
\end{minipage}
\end{figure}
\end{center}

\noindent  We now select three of the roots (eigenvalues),  $h = 366.6, ~326, ~\& ~290.2$ for $n$ = 10 and $h = 370.74, ~333.23 ~\& ~297.87$ for $n$ = 3 for which we compute the eigenstates and then, the perturbations, respectively. In Figures (\ref{v8}) and (\ref{v9}) we have shown the potential function for $n=10$ and $n=3$ spikes respectively. In our discussion later we will be working with the top three eigenvalues indicated by red lines. We show some of the eigenvalues and other parameters for spiky strings with $n=10$ and $n=3$ spikes in the table (\ref{T1}).
\begin{table}
\caption{Values of the parameters and eigenvalues for two different cases }\label{TT1}
\begin{center}

%\captionof{table}{}\label{TT1}
\begin{tabular}{|c|c|c|c|c|c|} \hline 
  $n$ & $k$ & $\alpha$ & $\rho_1$ & $\rho_0$ & $h$\\
  \hline
  & & & & & $9.74, ~18.09, ~27.53, ~40.07$ \\
   & & & & & $55.68, ~72.95, ~92.7, ~114.41$ \\
   $10$ & $0.985264$ & $1.00914$ & $3.18$ & $1.068$ & $-$ \\ 
  % & & & & & $-$ \\
    & & & & & $-$ \\
    & & & & & ${\bf 290.2},~{\bf 326},~{\bf 366.6}$ \\
    & & & & & $-$ \\
   % & & & & & $-$ \\
  %\, 27.53,\,40.07,\, .....,{\bf 290.2},\, {\bf 326}, \, {\bf 366.6},...$\\
  \hline
   & & & & & $7.50, ~15.23, ~24.58, ~36.65$ \\
    & & & & & $49.91, ~66.12, ~84.07, ~103.97$ \\
  $3$ & $0.985653$ & $1.01232$ & $2.645$ & $0.45$ & $ -$\\
   %& & & & & $-$ \\
    & & & & & $-$ \\
     & & & & & ${\bf 297.87}, ~\bf {333.23}, ~\bf {370.74}$ \\
     & & & & & $-$ \\
        % & & & & & $-$ \\
  \hline
 % $B,\, D$ & $\tau^2h$\\
\end{tabular}
\label{T1}
\end{center}
\end{table}

\noindent Let us now write down the general form of the perturbations.  For the $\tau$
part of the solution we take only the real part i.e $\cos \beta \tau$. So the general form of the perturbations are 
\begin{equation}
\begin{split}
\delta t &= \phi n^0 =  \frac{\epsilon ~\omega~ \cos \beta \tau ~\sqrt{\delta~ \gamma~ \sn^2 (u,k)- 2\delta~ (\delta+1) + \frac{\eta~ \delta}{\sn^2(u,k)}}}
{\alpha - \gamma~ \sn^2(u,k)}~\sn^2(u,k) ~\cn(u,k)\\ &  ~\dn(u,k) \sum_{m=0}^\infty C_m~\sn^{2m}(u,k) \, \, ,
\end{split}
\end{equation}
\begin{equation}
\begin{split}
\delta \rho & = \phi n^1 =  \frac{-\epsilon~\cos \beta \tau ~\sinh 2\rho_0}{\sqrt{\gamma^2~ \sn^4(u,k) - 2 \gamma~ (\delta+1)~ \sn^2(u,k) + \alpha~ \delta}} ~\sn^2(u,k) ~\cn(u,k) ~\dn(u,k) \\
&\sum_{m=0}^\infty C_m~\sn^{2m}(u,k) \, \, ,
\end{split}
\end{equation}
\begin{equation}
\begin{split}
\delta \theta & =  \phi n^2  =  \frac{\epsilon~\cos \beta \tau ~\sqrt{\delta~ \gamma~ \sn^2 (u,k)- 2\delta~ (\delta+1) + \frac{\eta~ \delta}{\sn^2(u,k)}}}
{\delta - \gamma~ \sn^2(u,k)} ~\sn^2(u,k) ~\cn(u,k) ~\dn(u,k) \\
& \sum_{m=0}^\infty C_m~\sn^{2m}(u,k) \, \, ,
\end{split}
\end{equation}
where the coefficients $C_m$ for different cases are shown in the table (\ref{T2}).
\begin{table}
\caption{Values of the coefficients $C_m$ (Eqn. \ref{s1}) for different cases}\label{TT2}
\begin{center}
\begin{tabular}{|c|c|c|c|}\hline
$n$ & $h$ & $\beta$ & $C_m$ \\
\hline
& 290.2 & $17.2$ & $C_0 = 1, ~C_1 = -28.03, ~C_2 = 248.6, ~C_3 = -910.09,$ \\
& & & $C_4 = 1259.1, ~C_5 = 95.33, ~C_6 = -952.5, ~C_7 = -657.46, ......$ \\
& & &  \\
10 & 326 & $18.22$ & $C_0 = 1, ~C_1 = -31.61, ~C_2 = 320.84, ~C_3 = -1390.1,$ \\
&  & & $C_4 = 2529.63, ~C_5 = -829.16, ~C_6 = -1947.05, ~C_7 = -236.62,.....$\\
& & & \\
& 366.6 & 19.32 & $C_0 = 1, ~C_1 = -35.67, ~C_2 = 413.83, ~C_3 = -2107.39,$ \\
& & & $C_4 = 4857.61, ~C_5 = -3524.75, ~C_6 = -2934.99, ~C_7 = 1953.34,......$\\
\hline
& 297.87 & 17.47 & $C_0 = 1, ~C_1 = -28.8, ~C_2 = 263.29, ~C_3 = -1001.62,$\\
& & & $C_4 = 1479.39, ~C_5 = -23.72, ~C_6 = -1145.42, ~C_7 = -644.47,.......$ \\
& & & \\
3 & 333.23 & 18.48 & $C_0 = 1, ~C_1 = -32.34, ~C_2 = 336.51, ~C_3 = -1503.37,$ \\
& & & $C_4 = 2865.34, ~C_5 = -1148.35, ~C_6 = -2160.91, ~C_7 = -16.34,......$ \\
& & & \\
& 370.74 & 19.49 & $C_0 = 1, ~C_1 = -36.09, ~C_2 = 423.95, ~C_3 = -2191.42,$\\
& & & $C_4 = 5157.14, ~C_5 = -3935.43, ~C_6 = -2986.85, ~C_7 = 2317.27,......$\\
\hline
\end{tabular}
\label{T2}
\end{center}
\end{table}\\

\noindent Figures (\ref{l1}), (\ref{l2}) \& (\ref{l3}) show perturbations $\delta t, ~\delta \rho, ~\& ~\delta \theta$ respectively for $n$ = 10 \& $n$ = 3 spikes and each for three different roots ($h$) i.e $h = 366.6, ~326 ~\& ~290.2$ for $n$ = 10 \& $h = 370.74, ~333.23 ~\& ~297.87$ for $n$ = 3 at constant $\tau$ (here $\tau = 0$) and we find
that {\em all perturbations are finite and oscillatory} in nature.
Thus, it is now clear that the solution of the DTV equation which
we have used leads to finite perturbations and therefore guarantees
the stability of the spiky strings. The stability is however 
apparent even without writing down the above expressions explicitly.
From the series solution given in Eqn.(\ref{s1}), the subsequent theorems
which involves positive powers of $\sn(u,k)$ and the expressions for
the normal components stated earlier, the finiteness of perturbations 
can indeed be qualitatively inferred.

%\begin{figure}[h]\label{f5}
 
%\end{figure}
%\end{center}
\subsubsection{Comments regarding the convergence issue}

\noindent Recall from the previous section that we have 
obtained the eigenfunctions by truncating the infinite series
involving powers of $\sn(u,k)$. Does the nature of the eigenfunctions
change if we include more terms in the series. It is necessary to check this aspect though it is not really related to the qualitative behaviour and finiteness of the solutions which is our primary concern here. 

\noindent Let us now address this convergence issue by considering more number of terms in the power series and observe the changes in the
behaviour of the eigensolutions. Figures (\ref{c1}), (\ref{c3}) \& (\ref{c4}) represent the perturbations $\delta t, ~\delta \rho ~\& ~\delta \theta$ respectively for $h=326$ and for four different cases: 1) effect of inclusion of the first five terms in the series shown in blue curve, 2) effect of including the first 21 terms of the series shown in the brown curve, 3) effect of including the first 33 terms represented by the green curve and 4) effect of including the first 43 terms represented by the red curve. Here one observes some interesting features. Inclusion of up to the first 21 terms does not quite change the qualitative nature of the curve. However, when we begin adding more terms its nature does gradually change and after including 33 terms the curve flips completely, as shown in the green curve. Thereafter, if one keeps adding  more number of terms in the series, we find that after adding 43 terms another overall sign flip occurs. The reason behind this flip can be easily found by considering the nature of the coefficients of the power series. We note that in the power series, the negative and positive terms  alternate at the beginning (i.e. for the first few terms). Thereafter the inclusion of terms till a certain order yields only positive terms. Subsequently, we have negative terms up to a further
order. This behaviour seems to repeat as we go on including more and more terms. When one considers first 33 terms one can find that there are sufficiently large number of negative coefficients so there is the flip in the curve.  Similarly when one considers 43 terms, one can see there are sufficiently large number of positive terms and there is another flip of sign.  Since this flipping occurs because of the alternating {\em negativity} or {\em positivity} of the coefficients one can ask which of two behaviours will eventually dominate if one includes sufficiently large number (ideally infinite) of terms.

\noindent Let us first check the case of {\em negativity}. Consider terms till the end of first set of negative values of the coefficients (i.e just before the set of positive coefficients begin). Thereafter, consider terms till the end of second set of negative values of the coefficients. A plot 
of the abovementioned two cases is shown in Figure (\ref{c2}) where the red curve represents the inclusions of 33 terms and the green curve 
includes 73 terms. One can easily see from Figure (\ref{c2}) that the amplitude of red curve is greater than the green one. Thus, one can say that the {\em negativity} effect is gradually diminishing. We have checked this by considering the inclusion of terms of very high orders and have found that the amplitudes of the negativity does indeed become smaller and smaller as more and more terms are included. It is therefore safe to state that when a very large number of terms are included the positive valued coefficients will dominate in the sum. This leads us to state that the nature of the perturbations will be very close to the 
curve represented in red in the Figure ({\ref{c2}). Hence the graphs shown earlier representing the perturbations do reflect the qualitative nature
of the perturbations correctly, though there will be quantitative differences (i.e in the exact values). Since our goal is to demonstrate that we have finite and oscillatory perturbations which imply the
stability of the string configuration, we believe that it is
not too crucial to be worried about exact numbers and values here.
 
%\newpage

\section{Concluding remarks}
In this article, which is a follow-up on our previous work \cite{bkp:2017}, we have demonstrated 
that the spiky string solutions in three dimensional anti-de Sitter spacetime are stable
against small deformations. Using the Jevicki-Jin embedding, we were able to reduce the
perturbation equation to an equation known in the literature as the Darboux-Treibich-Verdier (DTV) equation. Thereafter, we used the solution of the DTV equation as stated in \cite{ching}.  This solution involves an infinite series given in
terms of the squares of the Jacobi sine elliptic functions. The eigenvalues are 
hidden in infinite continued fractions. By numerically solving the infinite continued fractions,
we obtain the eigenvalues and the eigenfunctions. As shown in the previous section, the convergence of our solutions has been tested explicitly. Subsequently, we constructed the
deformations $\delta x^i$ and through plots, we have shown that for the given eigenvalues
the perturbations never diverge. We have checked our results for several eigenvalues.
Thus, we can safely conclude that the spiky string solutions are indeed stable. The perturbations (for specific eigenvalues), as obtained in our work, may also be used to study the normal mode fluctuations of the spiky string worldsheet embedded in $AdS_3$. 
Unlike our previous work, where we dealt with spiky strings in flat spacetime
(which do not have any link with AdS/CFT), here, it would certainly be
interesting to understand the role of the perturbations, in relation to the 
AdS/CFT correspondence. Furthermore, generalisations to higher dimensional AdS backgrounds
and a study of perturbations of the string configurations therein will be of interest in future work. The spectrum of perturbations may also be linked to the computation of quantum corrections \cite{tirziu} in some way. Finally, as an aside, the eigenvalues and eigenfunctions of the special
case of the DTV equation, which we have obtained by solving the continued fraction
numerically can be explored in the context of an analog quantum mechanics problem with a 
DTV potential, as well as its periodic generalisations. We hope to return to such
issues in future.

\section*{Acknowledgements}
\noindent We would like to thank Monodeep Chakraborty, Centre for Theoretical Studies,
IIT Kharagpur, India for his invaluable help in numerically solving the continued
fraction, which eventually led to a better, quantitative understanding of the problem. We would also like to thank Avinash Khare for useful discussions.

\end{document}